\documentclass[aps,pra,twocolumn,superscriptaddress]{revtex4-1}

\usepackage{amssymb,latexsym,amsmath,epstopdf,graphicx,color,natbib}

\def\e{\mathop{\rm \mbox{{\Large e}}}\nolimits}

\begin{document}
\title[Optical response of a hole pair]{The role of surface plasmon polaritons in the optical response of a hole pair}
\author{F. de Le\'on-P\'erez}
\affiliation{Centro Universitario de la Defensa de Zaragoza, Ctra. de Huesca s/n, E-50090 Zaragoza, Spain}
\affiliation{Instituto de Ciencia de Materiales de Arag\'on and Departamento de F\'isica de la Materia Condensada, CSIC-Universidad de Zaragoza, E-50009 Zaragoza, Spain}
\author{F.J. Garc\'ia-Vidal}
\affiliation{Departamento de F\'isica Te\'orica de la Materia Condensada, Universidad Aut\'onoma de Madrid, E-28049 Madrid, Spain}
\author{L. Mart\'in-Moreno}
\affiliation{Instituto de Ciencia de Materiales de Arag\'on and Departamento de F\'isica de la Materia Condensada, CSIC-Universidad de Zaragoza, E-50009 Zaragoza, Spain}

\begin{abstract} 
The optical emittance of a hole pair perforated in an opaque metal film is studied from first-principles using the coupled-mode method.  The geometrical simplicity of this system helps to understand the fundamental role played by surface plasmon polaritons (SPPs) in its optical response.  A SPP interference model without fitting parameters is developed from the rigorous solution of Maxwell's equations.  The calculations show that the interference pattern of the hole pair is determined by two scattering mechanisms: (i) re-illumination of the holes by the in-plane SPP radiation and (ii) an effective impedance depending on the single-hole response.  The conditions for constructive and destructive interference only depend on the phase difference provided by  each of the two scattering mechanisms.
\end{abstract}

\pacs{73.20.Mf, 78.67.-n, 41.20.Jb}

\maketitle

\section{Introduction}
The extraordinary transmission through nanohole arrays milled into metallic films \cite{EbbesenN98} is attributed to the resonant excitation of surface electromagnetic (EM) modes  by the incident light \cite{LMMPRL01}. In the optical regime, these surface EM modes are surface plasmon polaritons (SPPs), modified by the metal corrugation.  The light-SPP coupling is made possible by the additional grating momentum provided by the scattering of the incident light by the hole array. Nevertheless, interference of excited SPPs is set up even for two  interacting holes \cite{SoenninchensenAPL00,SchoutenPRL05, LalannePRL05,AlarverdyanNP07,PacificiOE08}. Increasing the number of holes, the transmission is enhanced due to better-defined peaks of the structure factor, appearing at the reciprocal
lattice vectors \cite{BravoPRL04}. It must also be noted that  light-SPP interaction is not the single mechanism behind the extraordinary optical transmission (EOT).  The EOT physical scenario is completed by the excitation of localized and Fabry-Perot modes \cite{WannemacherOC01,AbajoOE02,Degiron200461,FJPRL05,MolenPRB05,PopovAO05,FJPRB06,RindzeviciusJPC07}, which
may also contribute the whole process (see  \cite{FJRMP10} for a comprehensive review).

The aim of the present paper is to study the interference pattern of the simplest interacting system: a hole pair. Since the original proposition of the ``nanogolf'' effect by S\"onninchsen {\it et al.} \cite{SoenninchensenAPL00}, several groups have measured the optical interaction of two holes, see for example  \cite{AlarverdyanNP07,PacificiOE08,AlegretNJP08}.  These groups have used basic SPP resonant models in order to explain the characteristic optical transmittance of the hole dimer, which oscillates as function of the hole-hole distance, with period equal to the SPP wavelength. These approaches have in common that the relevant scattering channels are assumed {\it ad hoc}: only SPP scattering channels are included in the final optical response. 

In this paper we make no such assumption and solve Maxwell's equations from first-principles using a coupled-mode method (CMM) \cite{FJRMP10,FdLPNJP08}. We shall consider the two possible radiative channels: freely propagating light radiated out-of-plane into the far field, and SPP power scattered along the metal plane. The out-of-plane power $P_{\rm rad}$, normalized to the power incident on the hole area, gives the far field transmittance $T$. This is the quantity commonly used to characterize EOT. However, to the best  of our knowledge,  the in-plane SPP power $P_{spp}$ has not yet been measured for a hole pair.  We shall analyze the relevant scattering mechanisms for each radiative channel. Moreover, we shall derive, without fitting parameters, the conditions for constructive and destructive interference, hereafter conditions for interference  (CI), that explain experimental interference patterns \cite{AlarverdyanNP07,AlegretNJP08}.

The paper is organized as follows. In the next section we briefly review the CMM and give the expressions for  $P_{spp}$ and $P_{\rm rad}$.  The assumptions behind the CMM and some cumbersome mathematical formula are reported in the Appendix. For the sake of completeness, section \ref{sec:singlehole}  summaries  the emittance of a single hole. Section \ref{sec:holepair} discusses the optical response of the hole pair. A sub-section is devoted to clarify  the scattering mechanisms dominating the conditions for interference. At the end, we outline the main conclusions of the paper.

\begin{figure}
 \centering
\includegraphics[height=3cm]{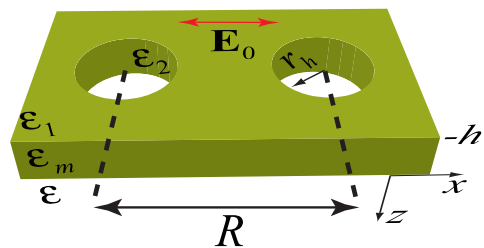}
\caption{(Color). Schematic representation of the hole-pair geometry.}
 \label{fig:scheme}
\end{figure}

\section{Theoretical framework}
\label{sec:me}

Figure \ref{fig:scheme} renders the hole-pair geometry studied in this paper. Two identical circular holes of radius $r_{h}$, separated by a distance $R$, are milled into an infinite metal film of thickness $h$ and dielectric function $\epsilon_{\rm m}$. In general, the metal film lays on a substrate with dielectric constant $\epsilon$, it is covered with a dielectric superstrate $\epsilon_1$, and the space inside the holes is characterized by a dielectric constant  $\epsilon_2$. For the sake of simplicity, $\epsilon=\epsilon_1=\epsilon_2=1$ is used along this paper. We consider in what follows that the metal film is illuminated by a normal-incident p-polarized plane wave, oriented along the main axis of the hole pair, as shown in Fig. \ref{fig:scheme}. We shall focus on the energy power radiated into the transmission region ($z>0$). 

Maxwell's equations are solved self-consistently using a convenient representation for the EM fields \cite{FJRMP10,FdLPNJP08}. In both substrate and superstrate the fields are expanded into an infinite set of plane waves with both p- and s-polarizations. Inside the holes the most natural basis is a set of circular waveguide modes. Convergence is fast achieved with a small number of such modes \cite{baudrionOE08,przybillaOE08}.  In fact, we shall see that the fundamental waveguide mode  is a good approximation for our problem. The assumptions behind this coupled-mode method, as well as its relevant constitutive quantities, are briefly review in the Appendix under the single mode approximation.
 
The flux power traversing the hole is distributed into  two channels \cite{FdLPNJP08}:  (i) out-of-plane radiation, freely propagating into the far-field, and (ii) SPP power, scattered  along the metal plane. The calculation of these two quantities is straightforward within the CMM after we know the amplitude of the fundamental waveguide mode at the hole openings $E'_i$, where $i=1,2$ labels each hole. For a normal-incident plane wave, both holes receive the same illumination $I$ (\ref{eq:I}), therefore $E'_1=E'_2 \equiv E'$ due to the symmetry of the system with respect to the central point of the hole pair; $E'$ hence reads
\begin{eqnarray}
\label{eq:Ep}
E'=\frac{G_{\rm \nu} I}{\left[G_{\rm sh}+G_{hh}(R)-\Sigma\right]^2-G^2_\nu},
\end{eqnarray}
where the hole-hole propagator $G_{hh}(R)$ (\ref{eq:green}) represents the coupling of the two holes as a function of the hole-hole distance $R$. This interaction can be seen as a re-illumination of the hole $i$ by the magnetic field  $G_{hh}E_j$ radiated from the other hole $j$. Notice that there is also a self-illumination term for each hole, $G_{\rm sh}$, which adds to the single-hole scattering mechanisms $\Sigma$ (\ref{eq:S}) and $G_{\rm \nu}$ (\ref{eq:Gnu}). Using  Eq. (\ref{eq:Ep}), the  out-of-plane power emitted by the hole pair simplifies to
\begin{eqnarray}
\label{eq:Prad}
  P_{\rm rad}(R)=\vert E' \vert^2 g_{\rm rad}(R),
\end{eqnarray}
where  the propagator $g_{\rm rad}(R)=g^{\rm sh}_{\rm rad}+g^{\rm int}_{\rm rad}(R)$ provides the far field radiated from the hole pair,  $g^{\rm sh}_{\rm rad}$ represents the contribution of each single hole,  and $g^{\rm int}_{\rm rad}(R)$ (\ref{eq:greenR}) is a term arising from the interference of the fields radiated by the two holes.  

On the other hand, we can obtain the power radiated into SPPs by computing the contribution from the plasmon pole in the propagator \cite{FdLPNJP08}.  The power of the scattered SPPs is first computed, at a point $r$ on the metal surface several SPP wavelengths away from the nearest edge of the hole pair, by integrating the in-plane radial component of the Poynting vector, defined with the SPP fields, on a cylindrical surface of radius $r$ and semi-infinity extension in $z>0$; the power in the plasmon wave is then calculated using the known decay length of the SPP. The integrated power reads
\begin{eqnarray}
\label{eq:Pspp}
 P_{spp}(R)= \vert E' \vert^2 g_{spp}(R),
\end{eqnarray}
where the propagator $g_{spp}(R)=g^{\rm sh}_{spp}+g^{\rm int}_{spp}(R)$ provides the total SPP  field radially scattered along all possible angular directions in the metal plane, $g^{\rm sh}_{spp}$ (\ref{eq:Gspp0}) represents the contribution of each single hole,  and $g^{\rm int}_{spp}(R)$ (\ref{eq:gspp})  is the interference term

The conservation of the energy flux for a lossless metal (\ref{eq:efc}), imposes a constrain to the real part of the full interaction propagator $G(R)=G_{\rm sh}+G_{hh}(R)$, which fulfills $\mbox{Re}[G(R)]=g_{\rm rad}+ g_{spp}$, see Appendix. We shall see that this relation is a good approximation for lossy metals at optical frequencies, and we shall use it in section \ref{sec:holepair}. However, the hole-hole interaction can not be fully understood without a previous knowledge of the optical response of a single hole,  which is briefly reviewed  in the next section.

\section{Single hole emittance}
\label{sec:singlehole}
\begin{figure}
 \centering
\includegraphics[height=6.5cm]{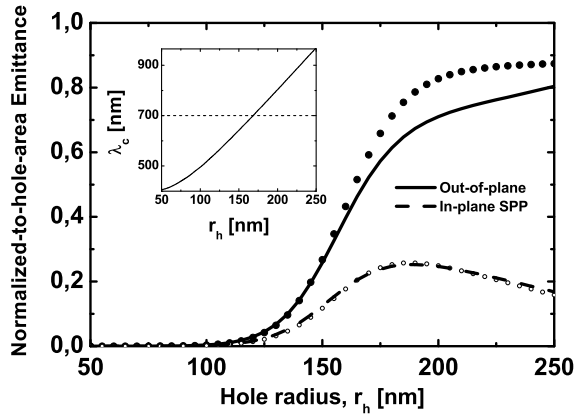}
\caption{Normalized-to-hole-area out-of-plane ($P_{\rm rad}$)  and in-plane SPP ($P_{spp}$) emittance as function of  the hole radius $r_{h}$ (in nm) for a single hole milled in a silver film, free standing on air ($h=250$ nm and $\lambda=700$ nm). Symbols and lines represent converged results  and the single mode approximation, respectively. The inset show the cutoff wavelength, $\lambda_{\rm c}$ (in nm).  }
 \label{fig:single_hole}
\end{figure}
The emittance spectrum of a single circular hole is described in this section for the sake of completeness, although this issue have been largely study, see for example \cite{SoenninchensenAPL00,WannemacherOC01,AbajoOE02,PopovAO05,FJPRB06,AlegretNJP08,FdLPNJP08,baudrionOE08,przybillaOE08,ChangOE05} and references therein. The behavior of both $P_{\rm rad}$ and $P_{spp}$ is depicted in Fig. \ref{fig:single_hole} as function of the hole radius, for a  free-standing Ag film with $h=250$ nm. The Ag dielectric function $\epsilon_{\rm m}(\lambda)$ is fitted to Palik's data \cite{Palik}. It is equal to $\epsilon_{\rm m}=-19.9+1.15 \: i$ for the incident wavelength, $\lambda=700$ nm, which is kept constant along the paper. Both  $P_{\rm rad}$ and $P_{spp}$ are normalized to the power incident on the hole area.

Fig. \ref{fig:single_hole}  renders $P_{spp}$ for both the fundamental mode approximation (dashed line) and converged results (open circles). Both curves practically overlap, so the fundamental mode is enough to achieve converged results for this emittance channel. For the out-of-plane emittance the agreement between single mode (solid line) and full calculations (full circles)  is slightly worse, but still the difference is less than 15\% and tendencies are well captured in the parameter range considered.

As already stressed in  Ref. \cite{baudrionOE08}, $P_{spp}(r_{h})$ presents a broad peak with maximum at $r_{h}=190$ nm, close to the cutoff radius, $r_{\rm c}=168$ nm for  $\lambda=700$ nm. The cutoff wavelength, $\lambda_{\rm c}$,  is represented in the inset of Fig. \ref{fig:single_hole} as a function of $r_{h}$. The resonance appears in the field at the opening $|E'|$ (not shown), while the decay for $r_{h}>r_{\rm c}$ is due to that in the single hole SPP propagator $g^{\rm sh}_{spp}$ (\ref{eq:Gspp0}).  For $r_{h}>r_{\rm c}$ most of the the energy is radiated out of the plane. In this case, both $g^{\rm sh}_{\rm rad}$ (not shown) and $P_{\rm rad}$ reach a fast saturation with the hole radius.  

\section{Optical response of a hole pair}

\label{sec:holepair}
We define the {\it normalized} hole pair emittance as the power radiated into each channel, out-of-plane (\ref{eq:Prad}) or in-plane-SPP (\ref{eq:Pspp}), divided by twice the corresponding emitted power of a single hole located at $R=0$, i.e.
\begin{eqnarray}
\label{eq:Jrad}
  \eta_{rad}(R)= \frac{ P_{\rm rad}(R)}{2 P^{\rm sh}_{\rm rad}}=|E'_N|^2  g^{\rm N}_{\rm rad}(R), \\
\label{eq:Jspp}
 \eta_{spp}(R)=\frac{P_{spp}(R)}{ 2 P^{\rm sh}_{spp}}=|E'_N|^ 2  g^{\rm N}_{spp}(R),  
\end{eqnarray}
where $E'_N$ is the ratio of the electric field at the hole openings 	
\begin{eqnarray}
\label{eq:rE}
E'_N (R) &=&  \frac{E'(R)}{E'_{\rm sh}} = \frac{\left[G_{\rm sh}-\Sigma\right]^2-G^2_\nu}{\left[G(R)-\Sigma\right]^2-G^2_\nu},
\end{eqnarray}
and we have used that the illumination of an isolated hole is equal to the illumination of each hole in the pair for a normal incident plane wave;
while the ratios for the out-of-plane and SPP propagators are given by
\begin{eqnarray}
\label{eq:rrad}
g^{\rm N}_{\rm rad}(R)= \frac{g_{\rm rad}(R)}{g^{\rm sh}_{\rm rad}}, \\
\label{eq:rspp}
 g^{\rm N}_{spp}(R) = \frac{g_{spp}(R)}{g^{\rm sh}_{spp}} .
\end{eqnarray} 

\begin{figure}
 \centering
\includegraphics[height=6.5cm]{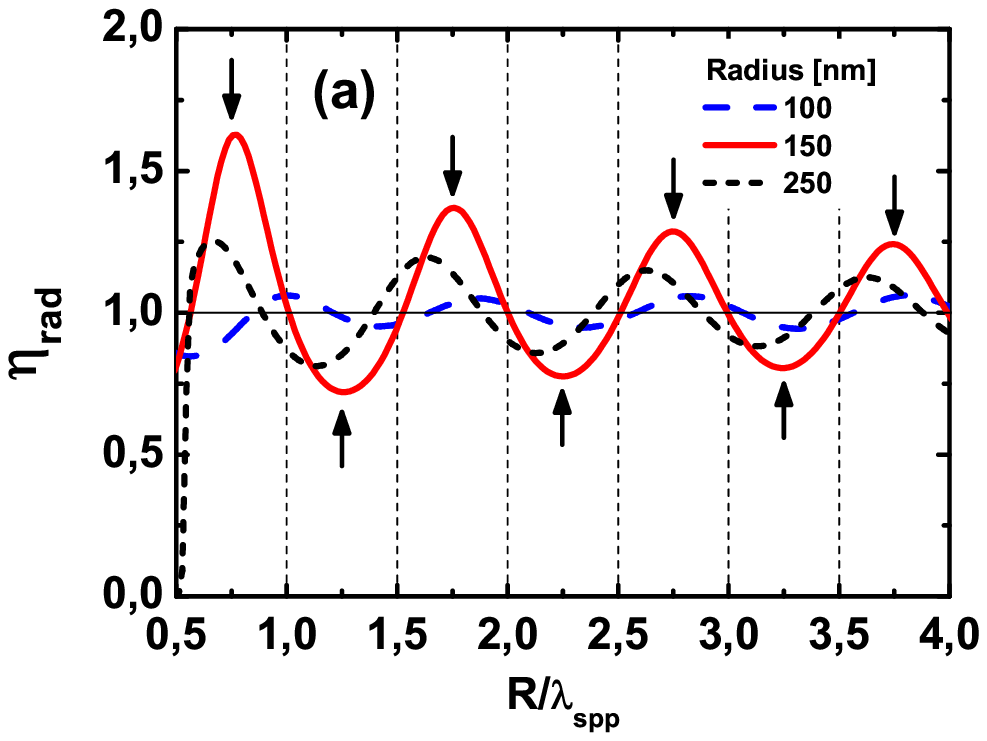}
\includegraphics[height=6.5cm]{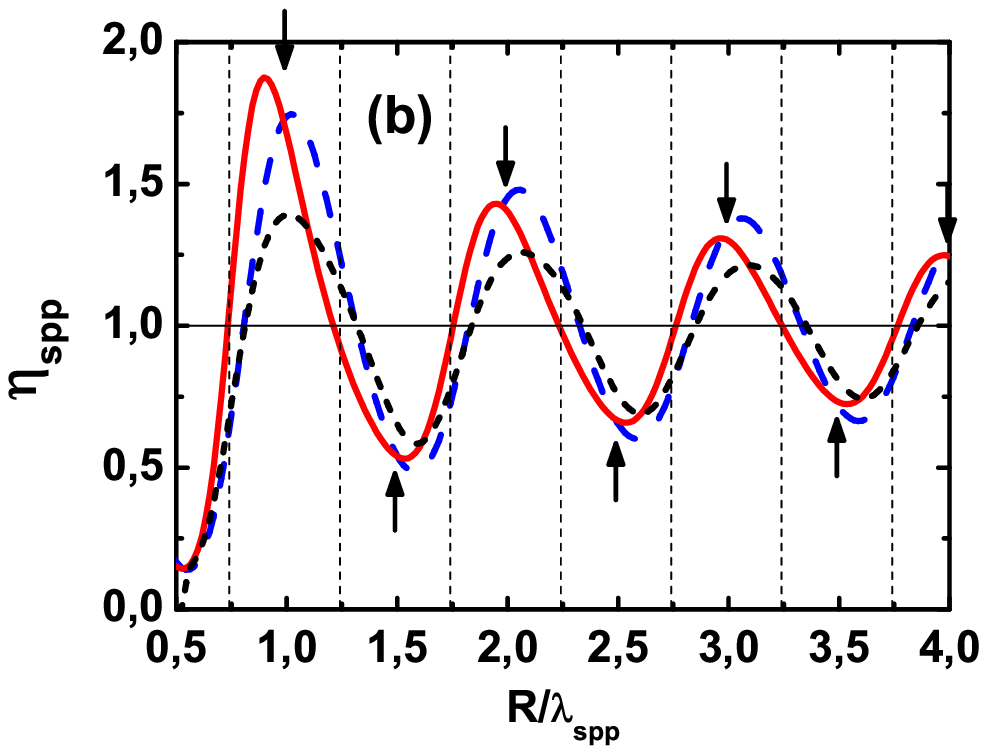}
\caption{(Color). Normalized out-of-plane emittance $\eta_{rad}$ (a) and normalized in-plane SPP emittance $\eta_{spp}$ (b) as function of $R/\lambda_{spp}$ for increasing $r_{h}$; $r_{h}=100$ nm (blue dashed line), 150 nm (red solid line), and 250 nm (black short-dashed line). The holes are milled in a free standing Ag film of thickness=250 nm. The illumination wavelength is $\lambda$=700 nm. The CI reported in \ref{sec:IC} are included in both (a) and (b) for $r_{h}=150$ nm: maxima (at $R/\lambda_{spp}=m-1/4$ for $\eta_{rad}$ and $R/\lambda_{spp} \approx m$ for  $ \eta_{spp}$) and minima (at $R/\lambda_{spp}=m+1/4$ for $\eta_{rad}$ and $R/\lambda_{spp} \approx m-1/2$ for  $ \eta_{spp}$) are represented with arrows, while vertical dashed lines are used for the condition $\eta=1$ (at $R/\lambda_{spp}=(m+1)/2$ for $\eta_{rad}$ and $R/\lambda_{spp} \approx (2m+1)/4$ for  $ \eta_{spp}$); $m=1,2,3,\dots$}
 \label{fig:Agh250r150}
\end{figure} 

The normalized emittances $\eta_{rad}$ and $ \eta_{spp}$  are depicted in Fig. \ref{fig:Agh250r150}  as a function of the hole-hole distance $R$ for increasing hole radius; $r_{h}=100$ nm (blue dashed line), 150 nm (red solid line), and 250 nm (black short-dashed line).  $R$ is normalized to the SPP wavelength $\lambda_{spp}=2\pi/\mbox{Re}[k_{spp}]$, where $k_{spp}$ (\ref{kspp}) is the SPP propagation constant in silver; $\lambda_{spp}=682.3$ nm for the chosen $\lambda=700$ nm.  

In accordance with experimental works \cite{AlarverdyanNP07,PacificiOE08}, the computed powers $\eta_{rad}$ and $ \eta_{spp}$  oscillate with period $\lambda_{spp}$. However, $\eta_{rad}$ behaves different than $ \eta_{spp}$ as a function of the hole radius. The amplitude of  $\eta_{rad}$ strongly oscillates with $r_{h}$. Indeed, increasing $r_{h}$ from 100 nm to 150 nm at fixed $R$	we can transform a maximum of $\eta_{rad}$ into a minimum. To the best our knowledge this dependence of $\eta_{rad}$ on $r_{h}$ has not been previously reported. Moreover, in the thin-film limit it has been found that the CI only depend on the edge-edge distance, and not on $r_{h}$. Further experimental work is needed to study the dependence on $r_{h}$ for opaque metal films and hole sizes larger that the metal skin depth (the region of the parameter space targeted in this paper). Nevertheless, it is worth stressing that the available experimental data  \cite{AlarverdyanNP07,PacificiOE08} report the same CI for very different systems \cite{note1}. 
In both cases $r_{h}$ is very small ($\sim \lambda/20$), but while  Ref. \cite{AlarverdyanNP07} considers a thin gold layer ($h=20$ nm) on a glass substrate, 
 Ref. \cite{PacificiOE08} uses an optically thick silver film, immersed in a medium with refractive index $n=1.45$. Both experimental CI are the same as for a third different system, the particular case $r_{h}=150$ nm in Fig. \ref{fig:Agh250r150} (a), i.e. maxima are at $R/\lambda_{spp}=m-1/4$, minima at $R/\lambda_{spp}=m+1/4$ (both represented with arrows), and $\eta_{rad}=1$ at $R/\lambda_{spp}=(m+1)/2$ (represented with vertical dashed lines), where $m=1,2,3,\dots$ 

In contrast, the amplitude of $ \eta_{spp}$ shows a stronger dependence on $R$, but does not present such large variations with size of the holes. Maxima of $ \eta_{spp}$  occur close to the  conditions for constructive interference of SPPs at the flat metal surface ($R=m\lambda_{spp}$), while minima appear close to conditions for destructive interference of SPPs between the holes ($R=(2m-1)\lambda_{spp}/2$). As the energy traversing the holes is distributed into the out-of-plane and in-plane channels (\ref{eq:efc}), we find in Fig. \ref{fig:Agh250r150} that $\eta_{rad}$ and $ \eta_{spp}$ behave as complementary scattering channels, with a relative contribution that changes as a function of both $R$ and $r_{h}$. Taking into account the interference pattern on the SPP channel we solve the apparent paradox  put forward in Ref. \cite{PacificiOE08}: although  $\eta_{rad}$ is described by an SPP interference model, there is neither a transmission enhancement nor suppression at the conditions for constructive interference  of  SPPs. 

\begin{figure}
 \centering
\includegraphics[height=6.5cm]{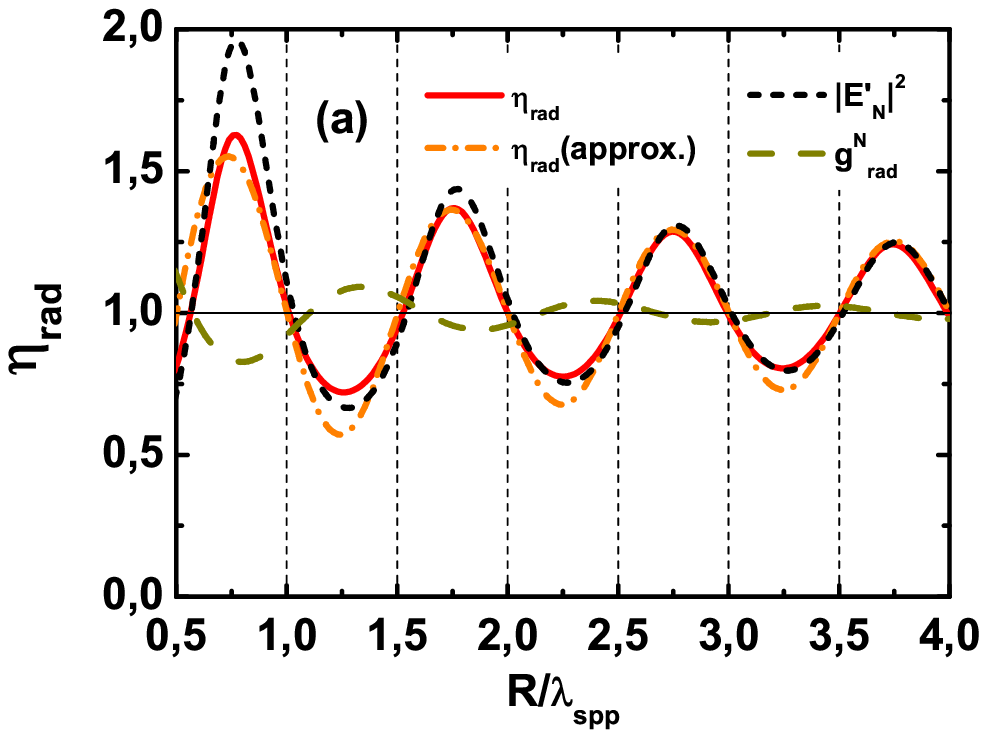}
\includegraphics[height=6.5cm]{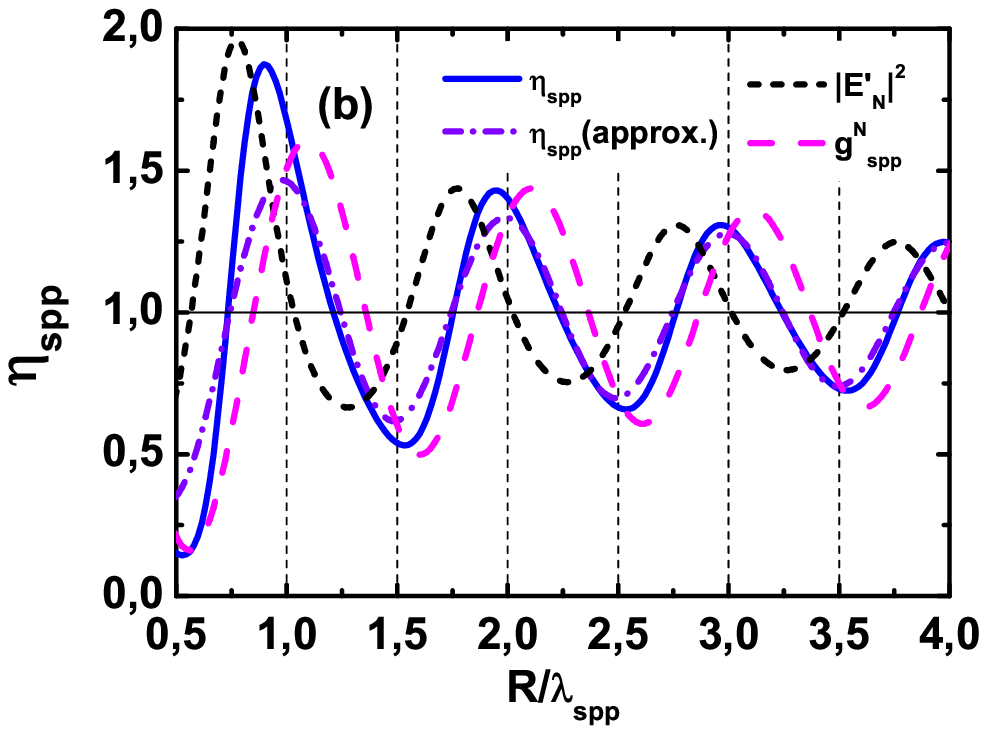}
\caption{(Color). (a) Normalized out-of-plane emittance $\eta_{rad}$ (red solid line), its constituent terms $|E'_N|^2$ (black short-dashed line) and $g^{\rm N}_{\rm rad}$ (dark-yellow dashed line), and the approximate expression for $\eta_{rad}$ (orange dash-dotted line) of  Eq. (\ref{eq:Pradapprox}). (b) Normalized SPP emittance $ \eta_{spp}$ (blue solid line), its constituent terms $|E'_N|^2$ (black short-dashed line) and $g^{\rm N}_{spp}$(magenta dashed line), and the approximate expression for $ \eta_{spp}$ (violet dash-dotted line) of  Eq. (\ref{eq:Psppapprox}). All these quantities are represented as function $R/\lambda_{spp}$. The hole radius is $r_{h}=150$ nm, the rest of parameters are the same as in Fig. \ref{fig:Agh250r150}.}
 \label{fig:j_vs_R}
\end{figure}

The out-of-plane radiation, $\eta_{rad}$ (\ref{eq:Jrad}), is mainly determined by $E'_N$, i.e. by the change in the field at the hole due to the presence of the other hole. This is illustrated in Fig. \ref{fig:j_vs_R}(a), where $\eta_{rad}$ is compared with both  $|E'_N|^2$ and $g^{\rm N}_{\rm rad}$. We observe that the interference between the radiative field of the two holes, given by $g^{\rm N}_{\rm rad}$,  practically does not change the total transmission for hole-hole distances larger than $2\lambda_{spp}$. Conversely, the normalized in-plane propagator $g^{\rm N}_{spp}$ plays an important role  setting up the CI for $ \eta_{spp}$ (\ref{eq:Jspp}), see Fig. \ref{fig:j_vs_R}(b). 
Although the contribution of $|E'_N|^2$ can not be neglected, the interference pattern of $g^{\rm N}_{spp}$ resembles the behavior of $ \eta_{spp}$. 

The CI developed in the next section strongly depend on the properties of the in-plane propagator $G_{hh}(R)$, which is behind the interference pattern of both radiative channels.  We use the following decomposition of the in-plane propagator
\begin{eqnarray}
\label{eq:Ghhdecomp}
 G_{hh}(R)=G^{hh}_{\rm rad}(R)+G^{hh}_{spp}(R)+G^{hh}_{\rm ev}(R),
\end{eqnarray}
where $G^{hh}_{\rm rad}(R)$ (\ref{eq:Gfp}) represents the contribution of radiative modes, $G^{hh}_{spp}(R)$ (\ref{eq:Gspphh}) designate the contribution of the plasmon pole to evanescent modes, and $G^{hh}_{\rm ev}(R)$ (\ref{eq:Grnp}) denotes the contribution of the remaining evanescent modes. This decomposition
is not only the most natural way of connecting $G_{hh}$ (\ref{eq:green}), to the radiative propagators $g_{\rm rad}$ (\ref{eq:Prad}) and $g_{spp}$ (\ref{eq:Pspp}), as well as to recover previous results for the PEC \cite{BravoPRL04}, it is also related to the decomposition proposed in  Ref. \cite{LalanneNP06} in order to compare SPP with non-SPP mediated interaction.

\begin{figure}
\centering
\includegraphics[height=6.5cm]{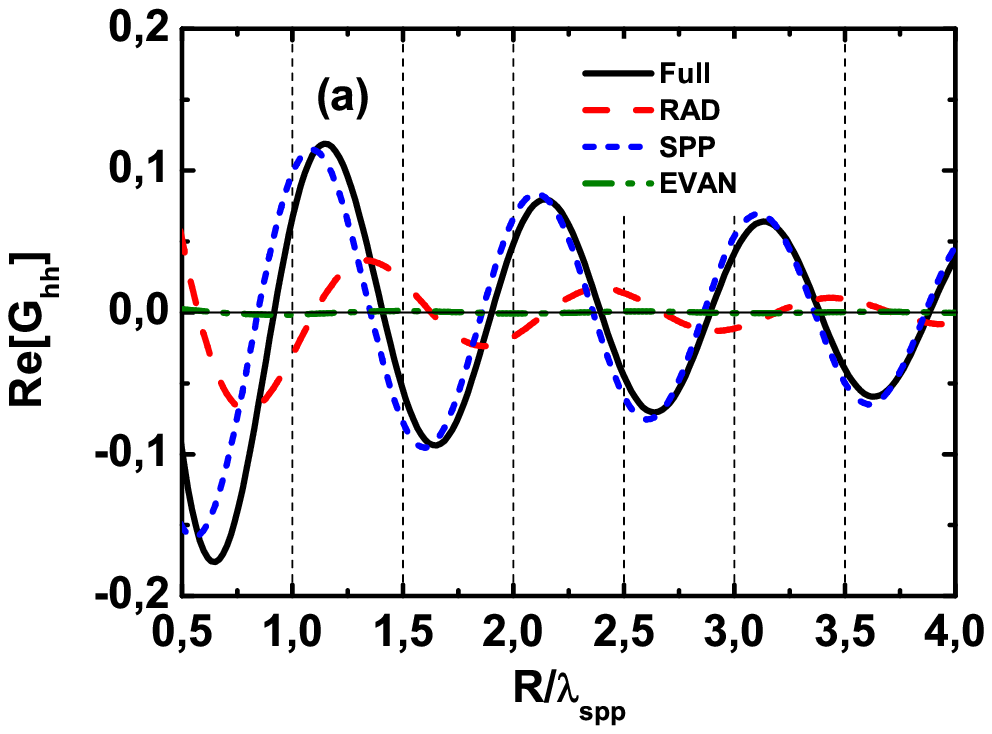}
\includegraphics[height=6.5cm]{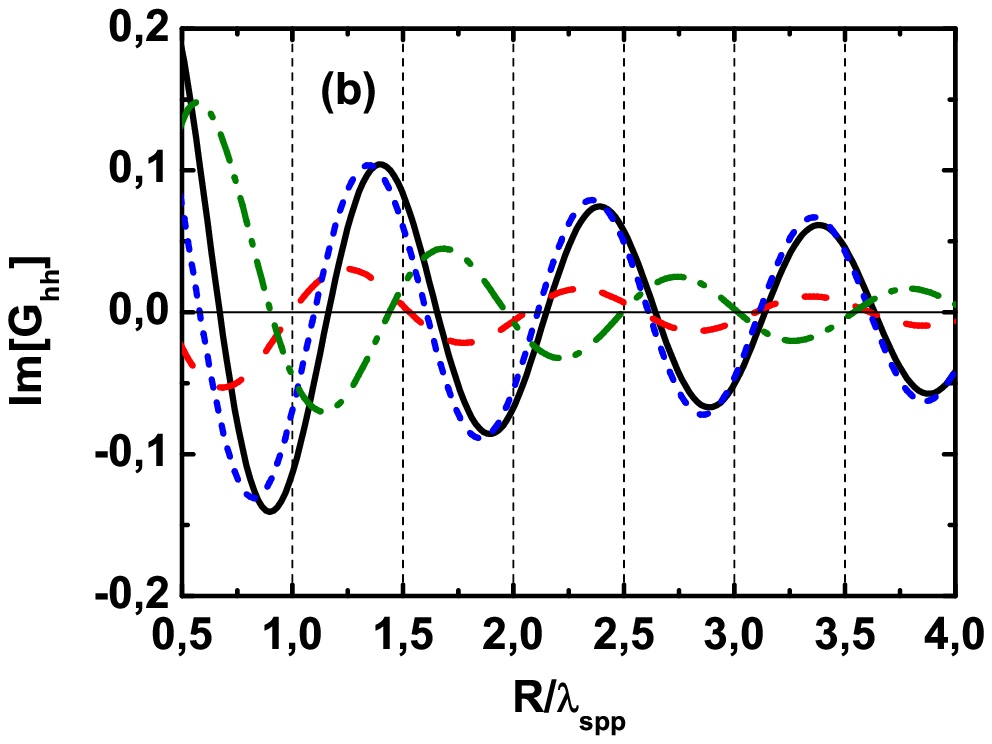}
\caption{(Color). Real (a) and imaginary (b) parts of the propagator $G_{hh}(R)$ (black solid line), as well as its constituent terms for radiative modes, $G^{hh}_{\rm rad}(R)$ (red dashed line), SPP modes, $G^{hh}_{spp}(R)$ (blue short-dashed line), and remaining evanescent modes $G^{hh}_{\rm ev}(R)$ (green dash-dotted line). We use the same geometrical parameters of Fig. \ref{fig:j_vs_R}.}
 \label{fig:green}
\end{figure}

The real and imaginary parts of   $G^{hh}_{\rm rad}(R)$, $G^{hh}_{\rm ev}(R)$, and $G^{hh}_{spp}(R)$ are compared with $G_{hh}(R)$ in Fig. \ref{fig:green} for the same parameters of Fig. \ref{fig:j_vs_R}. The most relevant feature observed in Figs. \ref{fig:green} (a) and (b) is that  the main contribution to $G_{hh}(R)$ comes from the SPP propagator, $G^{hh}_{spp}(R)$, which has a simple analytical form (\ref{eq:Gspphh}).  This allows us to find analytical expression for CI that will be presented in the next section.  Notice that the agreement between $G_{hh}(R)$ and $G^{hh}_{spp}(R)$ has been previously reported for 1D defects separated a distance larger that $2-3\lambda$ \cite{FLTPRB05}. Regarding non-SPP channels, $G^{hh}_{\rm rad}(R)$ decays faster than $G^{hh}_{spp}(R)$ being negligible small for $R$ equal to a few $\lambda_{spp}$. On other hand, the real part of $G^{hh}_{\rm ev}(R)$ is vanishing small (see Appendix), while its imaginary part is in anti-phase to $G^{hh}_{\rm rad}(R)$. It must be noted that, as expected, the relative contribution of the different propagators changes when we approach the PEC limit \cite{LalanneNP06,FLTPRB05,SondergaardPRB04,LiuN08}.
 
\subsection{Conditions for Interference}
\label{sec:IC}
In this section we compute the conditions for constructive and destructive interference of both out-of-plane and in-plane radiative powers. We start with the simpler of these two quantities, $\eta_{rad}$. Three simplifications help in finding the results for $\eta_{rad}$. First, its interference pattern is accurately described by the normalized square field amplitude, $\eta_{rad} \approx |E'_N|^{2}$, see \ref{fig:j_vs_R}(a). Second, $G_{hh}(R) \approx G^{hh}_{spp}(R)$, as we have learned from Fig. \ref{fig:green}. Third, $G^{hh}_{spp}(R) \ll G_{\rm sh}$. This last approximation is valid for $R \gg \lambda_{spp}$, but we shall see it gives results that work surprisingly well even for $R \sim \lambda_{spp}$. Expanding $E'_N$ (\ref{eq:rE}) into Mclaurin series of $G^{hh}_{spp}(R)/G_{\rm sh}$ and keeping only the leading term, we find
\begin{eqnarray}
\label{eq:Enapprox}
 E'_N =\frac{E'}{E'_{\rm sh}} \approx  1-2 \, Z_E \, G^{hh}_{spp}(R), \nonumber 
\end{eqnarray}
where $Z_E=E_{\rm sh}/I$ (\ref{eq:ZE}) is the effective impedance of a single hole, which gives the modal amplitude at the hole opening as a function of the illumination. 
From the simplified expression for $E'$ we can deduce that the interference pattern of the hole pair is set up by both the single hole impedance and the re-illumination of one hole by the other. 

The CI for $\eta_{rad}$ can be written in terms of the phase shift of both $Z_E$ and $G^{hh}_{spp}(R)$. We thus define the single-hole phase shift, $\phi_{ZE}$,  from  $Z_E=|Z_E| \exp(i\phi_{ZE})$, as well as the phase difference acquire by the SPP when traveling from one hole to the other, $\phi_{hh}$, from  $G^{hh}_{spp}(R)=|G^{hh}_{spp}(R)|\exp(i\phi_{hh})$. An approximate expression for the $\phi_{hh}$ can be obtained replacing the Hankel function in $G^{hh}_{spp}$ (\ref{eq:Gspphh}) by its asymptotic expression, $H'^{(1)}_1(x) \approx (\pi x/2)^{ -1/2}\exp[i (x-\pi/4)]$. We have then $\phi_{hh}=k_{spp}R-\pi/4$. Keeping again the leading term in the expansion of $|E'_N|^2$, we obtain \begin{eqnarray}
\label{eq:Pradapprox}
  \eta_{rad}\approx 1-4\left|Z_E G^{hh}_{spp} \right| \cos(k_{spp} R+\phi_{ZE}-\pi/4). 
\end{eqnarray}
This equation clearly shows that the out-of-plane radiation depends both on the optical path traveled by the SPP when going to one hole to the other and the phase picked up by the field given the extra illumination provided by the SPP coming from the other hole. 
The approximate equation (\ref{eq:Pradapprox}) is compared with full calculations in Fig. \ref{fig:j_vs_R}(a). We find that  Eq. (\ref{eq:Pradapprox}) slightly underestimates $|E'_N|^2$ for $R < \lambda_{spp}$, but the agreement is excellent for $R> \lambda_{spp}$.  This nice agreement is related to the fact that non-SPP waves decays faster than SPP waves as a function of the distance, see Fig. \ref{fig:green}. The leading role of SPP waves for large $R$ have been already stressed in   \cite{LalanneNP06,FLTPRB05,SondergaardPRB04,LiuN08}. 
Eq. (\ref{eq:Pradapprox}) also agrees with the one proposed in \cite{PacificiOE08} following an intuitive interference plasmon model, which, in contrast to first-principles derivation of  (\ref{eq:Pradapprox}), contains fitting parameters. 

It is straightforward to derive the CI of $\eta_{rad}$ from  Eq. (\ref{eq:Pradapprox}) assuming that the absolute value of $G^{hh}_{spp}$ changes smoothly  with $R$, and that the dependence on $R$ mainly comes from its phase. Then we have that extrema of $\eta_{rad}$ appear at  
\begin{eqnarray}
\label{eq:rEext}
 k_{spp}R-\frac{\pi}{4}+\phi_{ZE}=n \pi,
\end{eqnarray}
where the integer value of $n$ is equal to $n=2m-1$ for maxima, $n=2m$ for minima,  and $m=1,2,3,...$; while the condition for $\eta_{rad}=1$ is shifted in $\pi/2$ with respect to the previous expression, i.e.
\begin{eqnarray}
\label{eq:rEeq1}
 k_{spp}R-\frac{\pi}{4}+\phi_{ZE}=(m+\frac{1}{2}) \pi.
\end{eqnarray}

\begin{figure}
 \centering
\includegraphics[height=6.5cm]{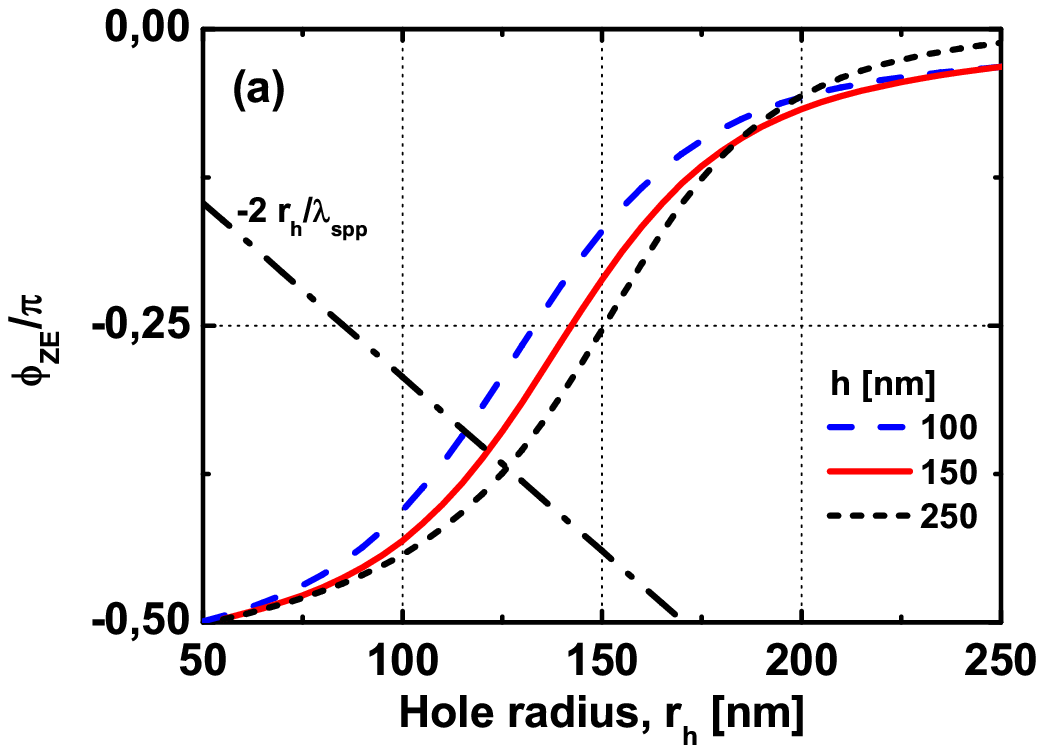}
\includegraphics[height=6.5cm]{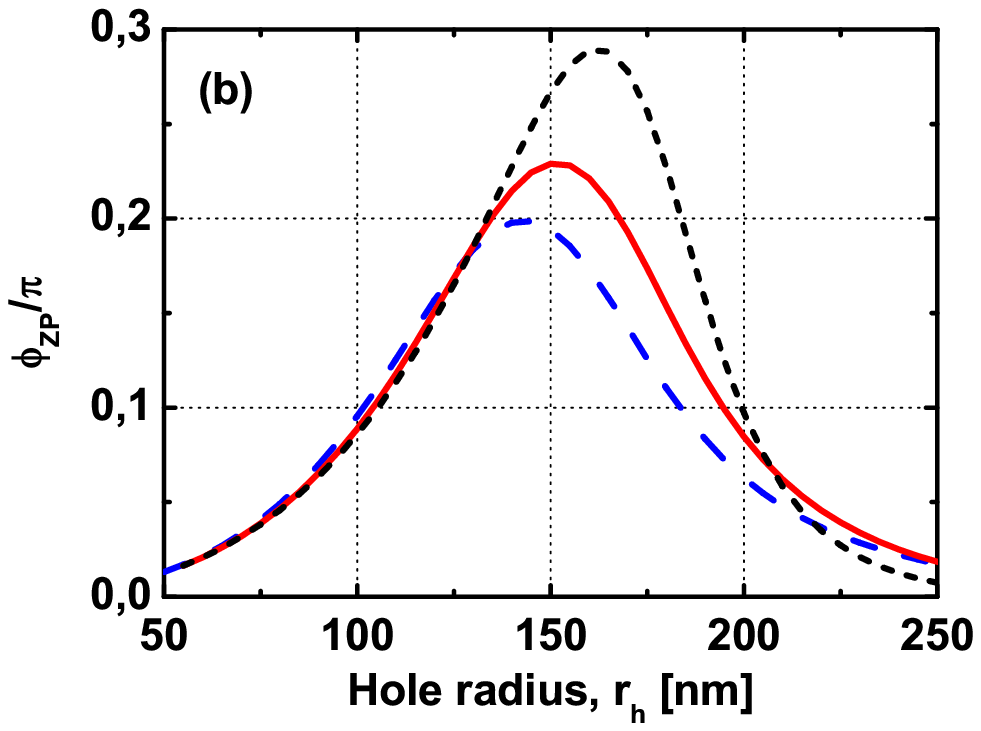}
\caption{(Color). (a) Single-hole phase shift for the out-of-plane emittance, $\phi_{ZE}$, as function of the hole radius, $r_{h}$, and for increasing metal thickness $h$; $h=100$ nm (blue dashed line), 150 nm (red solid line), and 250 nm (black short-dashed line). The dash-dotted  line represents the hole diameter normalized by $\lambda_{spp}$. (b) SPP phase shift for the in-plane emittance of a single hole, $\phi_{ZP}$.}
 \label{fig:phivsrh}
\end{figure}

The single-hole phase shift, $\phi_{ZE}$, is depicted in Fig. \ref{fig:phivsrh}(a) as function of the of the hole radius, $r_{h}$, and for increasing metal thickness; $h=100$ nm (blue dashed line), 150 nm (red solid line), and 250 nm (black short-dashed line). The large variation in $\phi_{ZE}$  as function of $r_{h}$ (up to $\pi/2$ for  increasing $r_{h}$ from 50 nm to 250 nm) accounts for the oscillations in $\eta_{rad}$ observed in Fig. 	\ref{fig:Agh250r150}(a). In Fig. \ref{fig:Agh250r150}(a) we compare full calculations with the CI given by  Eqs. (\ref{eq:rEext}) and (\ref{eq:rEeq1}) for the case $r_{h}=150$, $h=250$ nm, for which $\phi_{ZE}=-\pi/4$ (see Fig. \ref{fig:phivsrh}(a)). An excellent agreement is obtained even for small values of $R/\lambda_{spp}$. Notice that $\eta_{rad}$ is largely independent on the metal thickness $h$ (although it is computed for a given value of of $h$ in optically thick film) given that $\phi_{ZE}$ is practically independent on $h$, see Fig. \ref{fig:phivsrh}.

Notice that the CI represented by  Eqs. (\ref{eq:rEext}) and (\ref{eq:rEeq1}), which are valid for a wide range of hole sizes (larger than the metal skin depth) and opaque metal films, are expressed in terms of the distance between the centers of the holes. A previous work \cite{AlarverdyanNP07} suggested that, for thin-metal films and small hole sizes, the CI are a function of the edge-edge distance, independently from the hole radius. In our notation, this could only occurs if  $\phi_{ZE}+2r_{h}/\lambda_{spp}=0$ in  Eq. (\ref{eq:Pradapprox}). However, we observe in Fig. \ref{fig:phivsrh}(a) that $-2r_{h}/\lambda_{spp}$  (dash-dotted line)  is equal to $\phi_{ZE}$ only  for a small region of the parameter space. This novel behavior demands further experimental work  on opaque metal films and hole sizes larger than the metal skin depth.

 Similar CI can be developed for the in-plane scattered power $ \eta_{spp}$. As commented in the discussion of Fig. \ref{fig:j_vs_R}(b), both terms $|E'_N|^{2}$ and $g^{\rm N}_{spp}$ contribute to $\eta_{spp}$ in  Eq. \ref{eq:Jspp}. We take the approximate expression of $|E'_N|^{2}$ from  Eq. (\ref{eq:Pradapprox}) and use the asymptotic expression $g^{\rm N}_{spp}=1+2J'_{1}(k_{spp}R)$ found in the Appendix. We recall that the last relation is exact when absorption is neglected, but otherwise it is still a good approximation. Using again the asymptotic expansion of the Bessel function, the in-plane emittance is thus simplified to
\begin{eqnarray}
\label{eq:Psppapprox}
  \eta_{spp} \approx |E'|^2 \left[1+2\sqrt{\frac{2}{\pi k_{spp} R}}\cos(k_{spp}R-\pi/4) \right].
\end{eqnarray}
This equation tells us that, given the normalized amplitude of the electric field at the hole opening $E'_N$, the interference pattern of the in-plane scattering power is determined by the SPP optical path between the two holes. However, in order to quantify the CI of $ \eta_{spp}$, we should include the modulation of the field given by $|E'_N|^2$ (\ref{eq:Pradapprox}). Expanding the two terms in Eq. (\ref{eq:Psppapprox}) up to  the first order in $G^{hh}_{spp}/G_{\rm sh}$, $\eta_{spp}$ can be straightforwardly rewritten to 
\begin{eqnarray}
\label{eq:PsppIC}
  \eta_{spp} \approx 1+4\left|Z_{spp} G^{hh}_{spp} \right| \cos(k_{spp} R+\phi_{ZP}-\pi/4), 
\end{eqnarray}
where the effective impedance for the SPP channel $Z_{spp}=Z_E-(4|G_{spp}^{sh}|)^{-1}$ takes into account both the excitation of the EM field inside the hole, characterized by $Z_E$, and  the excitation of the SPP at the hole, given by $(4|G_{spp}^{sh}|)^{-1}$. Like for the out-of-plane channel, the approximate  Eq. (\ref{eq:PsppIC}) shows an excellent agreement with full calculations in Fig. \ref{fig:j_vs_R}(b).  However, the behavior of $\phi_{ZP}$ (defined from $Z_{spp}=|Z_{spp}|e^{i\phi_{ZP}}$)  differs from $\phi_{ZE}$. Fig. \ref{fig:phivsrh}(b) renders $\phi_{ZP}$ as a function of $r_{h}$, showing a characteristic peak centered near the cutoff radius $r_{\rm c}=168$ nm,  cf. Fig. \ref{fig:single_hole}. The phase difference with respect to $\phi_{ZE}$ is about $\pi/2$ for $r_{h} \leqslant r_{\rm c}$, and decreases to zero for $r_{h}>r_{\rm c}$. 

The extreme values of $ \eta_{spp}$  (\ref{eq:PsppIC}) satisfy
\begin{eqnarray}
\label{eq:rPext}
 k_{spp}R-\frac{\pi}{4}+\phi_{ZP}=n \pi,
\end{eqnarray}
where the integer value of $n$ is equal to $n=2m$ for maxima, $n=2m-1$ for minima, and $m=1,2,3,...$ Notice that the values of $n\pi$ for $ \eta_{spp}$ are shifted in $\pi$ with respect to the extreme values of $\eta_{rad}$ (maxima are replaced by minima, and vise versa). This shifting is determined by the fact that the power traversing the hole is radiated into two complementary channels: $\eta_{rad}$ and $ \eta_{spp}$. As for $\eta_{rad}$, the condition $ \eta_{spp}=1$ is shifted in $\pi/2$ with respect to the previous expression for extreme values, i.e.
\begin{eqnarray}
\label{eq:rPeq1}
 k_{spp}R-\frac{\pi}{4}+\phi_{ZP}=(n + \frac{1}{2})\pi.
\end{eqnarray}
In \ref{fig:Agh250r150}(b) we compare full calculations with the CI for the case $r_{h}=150$, $h=250$ nm, for which  $\phi_{ZP}=0.27 \; \mbox{rad} \approx \pi/4$ (see Fig. \ref{fig:phivsrh}(b)). As for $\phi_{ZE}$,  an excellent agreement is obtained.

\section{Conclusions}

We have studied the emission pattern of a hole pair, focusing our attention in the role played by SPP resonances.  Starting from the rigorous solution of the problem, we have developed a SPP interference model that does not contain fitting parameters. This model provides simple analytical expressions for the interference pattern of both the out-of-plane and in-plane radiation channels, which nicely agree with full calculations for noble metals at optical frequencies. 

In agreement with experimental reports, both radiated powers oscillate with period $\lambda_{spp}$. However, they show different trends as a function of the hole-hole distance and the hole radius. The amplitude of $\eta_{rad}$ strongly oscillates with the hole radius, while the amplitude of $\eta_{spp}$ has a stronger dependence on R, but does not present such large variations with the hole size. 

Maxima of $\eta_{spp}$  occur close to the  conditions for constructive interference of SPPs at the flat metal surface ($R=m\lambda_{spp}$), while minima appear close to conditions for destructive interference of SPPs between the holes ($R=(2m-1)\lambda_{spp}/2$). CI for $ \eta_{spp}$ are shifted in $\pi$ with respect to those of $\eta_{rad}$ (maxima are replaced by minima, and vise versa), because $\eta_{rad}$ and $ \eta_{spp}$ are two complementary channels. The power traversing the hole is distributed into these two channels.
 
We have also shown that  two scattering mechanisms determine the  interference pattern of  the hole pair: (i) re-illumination by the in-plane SPP radiation and (ii) an effective impedance depending on the single-hole response.  The conditions for interference only depend  on the phase difference provided by  each of the two scattering mechanisms. The large variation in the effective impedance of the single hole accounts for the oscillations of $\eta_{rad}$ as a function of the hole size.
 
\begin{acknowledgments}
The authors gratefully acknowledge financial support from the Spanish Ministry of Science and
Innovation under grants MAT2009-06609-C02, CSD2007-046-NanoLight.es, and AT2009-0027.
\end{acknowledgments}

\appendix

\section{Coupled-mode method}
\label{app:me}
In this section we briefly review the coupled-mode method for the optical transmission through holes, under the fundamental waveguide mode (TE$_{11}$) approximation. We refer to   \cite{FdLPNJP08} for the expressions of the full multimode formalism and their derivation. Within the CMM, Maxwell's equations are solved self-consistently using a convenient representation for the EM fields. In both substrate and superstrate (see Fig. \ref{fig:scheme}), the fields are expanded into an infinite set of plane waves with both p- and s-polarizations. Inside the holes the most natural basis is a set of circular waveguide modes \cite{stratton}.  The parallel components of the fields are matched at the metal/dielectric interface using surface impedance boundary conditions (SIBCs) \cite{jackson}. Although SIBCs neglect the tunneling of EM energy between the two metal surfaces, this effect is not relevant for a metal thickness larger than a few skin depths.

At the lateral walls of the holes we choose the PEC approximation for the sake of analytical simplicity. We are thus neglecting absorption losses at the walls. Nevertheless, we upgrade the PEC approximation introducing two phenomenological corrections. First, the propagation constant of the PEC fundamental mode is replaced by the one computed for a real metal. This improves the comparison between CMM and both experimental and FDTD results for both the spectral position of the peaks and the dependence of optical properties on the metal thickness. Second, enlarging the radius of the hole by one
skin depth simulates the real penetration of in field into the metallic walls. This value for the enlargement provides the best agreement
with FDTD simulations for an infinite periodic array of
holes \cite{przybillaOE08}.

After matching the fields at the interface  we arrive to the following system of tight binding-like equations
\begin{eqnarray}
[ G_{\rm sh}-\Sigma ] E_1(R)+G_{hh}(R) E_2(R)-G_{\rm \nu} E'_1(R) = I_1 , \nonumber \\
\left[ G_{\rm sh}-\Sigma \right]  E'_1(R)+G_{hh}(R) E'_2(R)-G_{\rm \nu} E'_2(R) = 0 ,\nonumber
\end{eqnarray}
where $E_i$ is the modal amplitude of the electric field at the input opening of the $i^{th}$ hole, $i=1,2$, and $E'_i$ is the same quantity but at the output opening. Two additional equivalent equations are needed for $E_2$ and $E'_2$. Other relevant quantity is the illumination provided by the normal-incident p-polarized  plane wave, with wavenumber $k_{\lambda}=2\pi/\lambda$ and admittance $Y_0=\sqrt{\epsilon_1}$, onto the lowest-energy mode
\begin{equation}
\label{eq:I}
I \equiv I_1=I_2=\frac{\sqrt{2 Y_0}}{1+z_{s} Y_0 }\frac{k_{\lambda}	}{\sqrt{u^2-1}},
\end{equation}
where  $z_{s}=\epsilon^{-1/2}_m$ is the metal impedance. In order to obtain a transmittance normalized by the flux impinging on the area covered by the holes, the illumination term $I$ already contains a factor $(\pi r^2_h Y_0)^{-1/2}$.  The constant $u$ satisfies $J'_1(u)=0$ \cite{stratton}, where $J(x)$ is the Bessel function of order $1$, and the prime denotes derivation with respect to its argument 

The quantities $\Sigma$ and $G_{\rm \nu}$ represent scattering mechanisms already present in single holes. $\Sigma$ is
related to the bouncing back and forth of the waveguide fields
inside the holes. Its value is
\begin{equation}
\label{eq:S}
 \Sigma= Y_w \frac{ f^+_w \e^{i k_zh}+f^-_w \e^{-i k_zh}}{ {f^+_w}^2 \e^{i k_zh}-{f^-_w}^2 \e^{-i k_zh}},
\end{equation}
where $k_z$ is the propagation constant of the waveguide mode, $h$ is the metal thickness, $f^{\pm}_{\rm w} = 1 \pm z_{s} Y_w$, $Y_w=k_z/k_{\epsilon_2}$ is the admittance for the excited TE$_{11}$ mode, and $k_{\epsilon_2}=\sqrt{\epsilon_2}k_{\lambda}$. The quantity 
\begin{equation}
\label{eq:Gnu}
 G_{\rm \nu}= 2 Y_w \left[
{f^+_w}^2 \e^{i k_z
h}-{f^-_w}^2\e^{-i k_zh} \right]^{-1} 
\end{equation}
reflects the coupling
between EM fields at the two sides of a given hole \cite{FdLPNJP08}. 

The propagator $G_{hh}(R)$ represents the coupling of the two holes. It results from the projection of the Green's dyadic onto the waveguide modes in the holes. For the TE$_{11}$ mode, $G_{hh}$ can be written as the following integral in the plane of the reciprocal space parallel to the metal surface
\begin{eqnarray}
\label{eq:green}
G_{hh}(R)&=&G_0\int^{\infty}_0  \left( \frac{G_{p}(q)}{q_{z}+z'_{s} }   + \frac{G_{s}(q)}{q_{z}^{-1} +z'_{s}} \right) q d q,
\end{eqnarray}
where $G_0 = 4 k^2_{\epsilon} r^2_h \sqrt{\epsilon}/(u^2-1)$ and the two terms in the integrand represent the contribution of p- and s-polarized plane waves in the infinite semi-space in contact with the metal surface. The denominators of these two terms stand for the response of the metal plane. In particular, the p-term has a pole at the SPP wavevector. The numerators $G_{p}$ and $G_{s}$ account for both the single hole response, which is a function of the hole radius $r_{h}$, and the hole-hole interaction, a function of $R$. They read 
\begin{eqnarray}
\label{eq:Gp}
 G_{p} (q,r_{h},R)=   \frac{ J^2_1 (k_{\epsilon} q r_{h})}{k^2_{\epsilon} q^2 r^2_h} J'_1 (k_\epsilon q R), \\
\label{eq:Gs}
G_{s} (q,r_{h},R)= \frac{J^{' 2}_1 (k_{\epsilon} q r_{h})}{\left( 1-\frac{k^2_{\epsilon} q^2 r^2_h}{u^2}\right)^2 } \frac{J_1(k_{\epsilon} q R)}{k_{\epsilon} q R}.
\end{eqnarray}	
The integrand is written in adimensional units normalizing the wavevector by $k_{\epsilon}=k_{\lambda} \sqrt{\epsilon}$. Notice that the $R$-dependent Bessel functions are obtained after the angular integration in the $\mathbf{k}_\parallel=k_\epsilon q (\cos \theta,\sin \theta)$ plane, where $\theta$ defines the direction of the component of wavevector parallel to the metal plane, $\mathbf{k}_\parallel$. The dielectric constant $\epsilon$ characterizes
the dielectric material in contact with the metal surface (see Fig. \ref{fig:scheme}, $\epsilon=\epsilon_1=1$ is used along this paper). 

The self-interaction term $G_{\rm sh}$ is obtained after tacking the limit $R \rightarrow 0$ in $G_{hh}(R)$, i.e. using the identities $\lim_{x \rightarrow 0} J'_1(x)=\lim_{x \rightarrow 0} J_1(x)/x=1/2$.  

Other relevant function is the effective impedance $Z_E$, which is determined by the the three scattering mechanisms of the single hole ($G_{\rm sh}$, $G_{\rm \nu}$, and $\Sigma$),
\begin{eqnarray}
\label{eq:ZE}
 Z_E=\frac{E_{\rm sh}}{I}=\frac{G_{\rm sh}-\Sigma}{(G_{\rm sh}-\Sigma)^2-G^2_\nu}. 
\end{eqnarray}

We compute $G_{hh}$ using the decomposition $G_{hh}=G^{hh}_{\rm rad}+G^{hh}_{spp}+G^{hh}_{\rm ev}$ (\ref{eq:Ghhdecomp}), where $G^{hh}_{\rm rad}(R)$  represents the contribution of radiative modes, $G^{hh}_{spp}(R)$  designate the contribution of the plasmon pole to evanescent modes, and $G^{hh}_{\rm ev}(R)$ denotes the contribution of the remaining evanescent modes.  The contribution of $G^{hh}_{\rm rad}(R)$ can be written in terms of the functions $g^{\rm int}_{\rm rad}(R)$ and $\Delta G^{hh}_{\rm rad}(R)$, which always take real values,
\begin{eqnarray}
\label{eq:Gfp}
 G^{hh}_{\rm rad}(R)=g^{\rm int}_{\rm rad}(R)+z'^*_s \Delta G^{hh}_{\rm rad}(R), \nonumber
\end{eqnarray}
where
\begin{eqnarray}
\label{eq:greenR}
g^{\rm int}_{\rm rad}(R)&=&G_0\int^1_0  d q  \left( \frac{q q_{z} G_{p}}{\vert q_{z}+z'_{s} \vert^2}  + \frac{q q_{z} G_{s}}{\vert 1+q_{z} z'_{s} \vert^2}  \right) 
\end{eqnarray}
provides the interference term of the far field radiated from the holes (\ref{eq:Prad}), while the term proportional to the metal impedance $z'_{s}$ reads 
 \begin{eqnarray}
\Delta G^{hh}_{\rm rad}(R)&=&G_0 \int^1_0  q  d q \left( \frac{G_{p}}{\vert q_{z}+z'_{s} \vert^2}  + \frac{q^2_z G_{s}}{\vert 1+q_{z} z'_{s} \vert^2}  \right). \nonumber
\end{eqnarray}
Both integrals are computed for free-propagating states ($0 \leq q \leq 1$). The real part of $z'_{s}$ is very small for typical noble metals, making $\mbox{Re}[G^{hh}_{\rm rad}] \approx g^{\rm int}_{\rm rad}$ and $\mbox{Im}[G^{hh}_{\rm rad}] \approx |z_{s}| \Delta G^{hh}_{\rm rad}$ a good approximation for $G^{hh}_{\rm rad}$. The same relations hold for the single hole propagator, $G^{sh}_{rad}$.

For non-propagating states ($q>1$) the integrand in $G_{hh}(R)$ is prolonged  into the complex $q$-plane, see  \cite{FdLPNJP08} for details. The residue of the Cauchy integral gives the SPP wave confined to the metal/air interface
\begin{eqnarray}
\label{eq:Gspphh}
 G^{hh}_{spp}=\pi i z'_{s} G_0 \frac{ J^2_1 ( k_{spp}r_{h})}{k^2_{spp} r^2_h} H'^{(1)}_1( k_{spp} R), 
\end{eqnarray}
where $k_{spp}$  is the parallel component of the SPP wavevector 
\begin{equation}
\label{kspp}
k_{spp} =k_\epsilon \left( \frac{\epsilon \; \epsilon_{\rm m}(\lambda)}{\epsilon_{\rm m}(\lambda)+\epsilon}\right)^{1/2} .
\end{equation}
Eq. (\ref{kspp}) defines $k_{spp}$ for a real metal and rigorous boundary condition at the metal/dielectric interface. In order to improve the accuracy our model we use  Eq. (\ref{kspp}) instead of the approximate SPP wavevector  for SIBCs, $k^{\rm sibc}_{spp} =k_\epsilon [ \epsilon( 1-\epsilon^{-1}_{\rm m}) ]^{1/2}$. Besides the coupling propagator $G^{hh}_{spp}$, we define the radiative propagator $g_{spp}$, which provides the total SPP  field radially scattered along all possible angular directions in the metal plane (\ref{eq:Pspp}). We have that $g_{spp}=g^{\rm sh}_{spp}+g^{\rm int}_{spp}$, where the single-hole contribution read
\begin{eqnarray}
\label{eq:Gspp0}
g^{\rm sh}_{spp}&=& \frac{\pi \vert z'_{s}\vert a_{\rm l} G_0}{2}  \left| \frac{ J_1 (k_{spp} r_{h})}{k_{spp} r_{h}} \right|^2, 
\end{eqnarray}
and the interference term is equal to
\begin{eqnarray}
\label{eq:gspp}
g^{\rm int}_{spp}(R)= g^{\rm sh}_{spp} \left[ 2 \mbox{Re}\left[J'_1(k_{spp}R)\right]+ J'_1(2 i \mbox{Im}[k_{spp}R])-\frac{1}{2} \right] , \nonumber \\
\end{eqnarray}
while $a_{\rm l} = |k_{\rm zp}| \mbox{Re}[k_{spp}] /(\mbox{Im}[k_{\rm zp}] |k_{spp}|)$, and $k_{\rm zp}=(k^2_{\rm \epsilon}-k^2_{spp})^{1/2}$.  Notice  that  $g^{\rm N}_{spp}=g^{\rm int}_{spp}/g^{\rm sh}_{spp}$ (\ref{eq:rspp}) is independent of $r_{h}$.

For the sake of convenience, the integral for non-SPP evanescent states is computed along the vertical contour  $q=1+ih \equiv q_+$, $h\in[0,\infty)$, after the integral variable  is changed from $q$ to $h$ 
\begin{eqnarray}
\label{eq:Grnp}
%\fl
 G^{hh}_{\rm ev}=\frac{G_0}{2} \int^\infty_0  q_+ d h \left(  \frac{G_{p}}{\kappa_{z}-i z'_{s}}-\frac{G^*_{p}}{\kappa^*_z-i z'_{s}}  \right. \nonumber \\
\left. -\frac{\kappa_{z} G_{s}}{1+i z'_{s} \kappa_{z}}+  \frac{\kappa^*_z  G^*_{s}}{1+i z'_{s} \kappa^*_z}\right),
\end{eqnarray}
 where
$\kappa_{z}=\sqrt{2i h-h^2}$, and the Bessel function $J_1(x)$ in both $G_{p}$ (\ref{eq:Gp}) and $G_{s}$ (\ref{eq:Gs}) is replaced by a Hankel function of the first kind $H^{(1)}_1(x)$. 

The propagator $G_{hh}$ is further simplified when the metal absorption is neglected, i.e. for $\mbox{Im}[\epsilon_{\rm m}]=\mbox{Re}[z'_{s}]=0$ and $\mbox{Im}[z'_{s}]=-|z'_{s}|$. We find for the radiative modes that $\mbox{Re}[G^{hh}_{\rm rad}] = g^{\rm int}_{\rm rad}$, and $\mbox{Im}[G^{hh}_{\rm rad}] = |z_{s}| \Delta G^{hh}_{\rm rad}$, while for SPP modes $\mbox{Re}[G_{spp}(R)]=g_{spp}(R)=g^{\rm sh}_{spp}[1+2J'_1(k_{spp}R)]$. 
For the remaining evanescent modes we obtain that $G^{hh}_{\rm ev}$ is a pure imaginary function. The same relations that hold for $G_{hh}$ are valid for  $G_{\rm sh}$. Therefore, only the radiative and SPP terms contribute to the real part of in-plane propagator,
\begin{eqnarray}
\label{eq:ReG}
\mbox{Re}[G_{\rm sh}+G_{hh}(R)]=g_{\rm rad}(R)+g_{spp}(R). 
\end{eqnarray}
Under the lossless metal approximation, the total power traversing the two holes simplifies to $P_{\rm hole}=\mbox{Re}[G_{\rm \nu} E E'^*]$  \cite{FdLPNJP08}. We rewrite it in terms of $E'$ with help of the relation $G_{\rm \nu} E=\left[G_{\rm sh}+G_{hh}(R)-\Sigma\right] E'$, i.e.
\begin{eqnarray}
\label{eq:Phole}
 P_{\rm hole}=\vert E' \vert^2 \mbox{Re}[G_{\rm sh}+G_{hh}(R)]. \nonumber
\end{eqnarray}
As $\Sigma$ (\ref{eq:S}) is purely imaginary for a lossless media, this term does not contribute to $P_{\rm hole}$. Using  (\ref{eq:ReG}), we then have 
\begin{eqnarray}
\label{eq:efc}
 P_{\rm hole}= P_{\rm rad}+P_{spp}.
\end{eqnarray}

This equality represents the conservation of the power flux traversing the hole. These results can be easily generalized to an arbitrary number of holes, waveguide modes, and non-cylindrical geometries. It is also worth to mention that including absorption the computed powers differ in less than 5\% from the lossless case, even for a large number of defects \cite{FdLPNJP08,baudrionOE08,przybillaOE08}. 

Finally, we recall that PEC is a particular case of a lossless metal with $z'_{s}=0$. In this case $G^{hh}_{\rm rad}=g^{\rm int}_{\rm rad}$, while for non-propagating states of a PEC only $G^{hh}_{\rm ev}$ survives because $G^{hh}_{spp}=0$ \cite{BravoPRL04}.

%\section*{References}
%\bibliographystyle{/Users/fernandodeleon-perez/unizar/linux/tex/njpstyle4}
%\bibliographystyle{njpstyle4}
%\bibliographystyle{njpstyle4.bst}
%\bibliographystyle{jphysicsB.bst}
%\bibliographystyle{unsrt} 
%\bibliography{/Users/fernandodeleon-perez/unizar/linux/tex/references}
%\bibliographystyle{apsrev4-1.bst} 
%\bibliography{references}
%\bibliography{/Users/fernandodeleon-perez/unizar/linux/tex/refbibdesk}

\begin{thebibliography}{30}
\bibitem [{not()}]{note1}%
  \BibitemOpen
  \href@noop {} {}\bibinfo {note} {The conditions for interference of Ref.
  \cite{AlarverdyanNP07} are given as function of the edge-edge distance. In
  order to rewrite it as a function of $R$ one should take into account that
  $2r_h \approx \lambda_{spp}/4$ for the measured $r_h$ and
  wavelength.}\BibitemShut {Stop}%
\end{thebibliography}
%

\end{document}